\documentclass{JAC2003}
\usepackage{graphicx}
\usepackage{booktabs}

\begin{document}
\title{THIRD-ORDER APOCHROMATIC DRIFT-QUADRUPOLE BEAMLINE\vspace{-0.5cm}}
\author{V.Balandin, 
R.Brinkmann, W.Decking, N.Golubeva\thanks{nina.golubeva@desy.de} \\
DESY, Hamburg, Germany}

\maketitle

\begin{abstract}
\vspace{-0.1cm}
In this paper we present the design of a straight drift-quadrupole system 
which can transport certain beam ellipses (apochromatic beam ellipses) 
without influence of the second and of the third order chromatic and geometric 
aberrations of the beamline transfer map.
\end{abstract}

\vspace{-0.15cm}
\section{INTRODUCTION}

\vspace{-0.1cm}
A straight drift-quadrupole system can not be designed in such a way that  
a particle transport through it will not depend on the difference 
in particle energies and this dependence can not be removed even 
in first order with respect to the energy deviations.
Nevertheless, the situation will change if instead of 
comparing the dynamics of individual particles  
one will compare the results of tracking of
monoenergetic particle ensembles through the system or
will look at chromatic distortions 
of the betatron functions appearing after their transport through the system.
From this point of view, as it was proven in ~\cite{ApochromIPAC10},
for every drift-quadrupole system 
there exists an unique set of Twiss parameters 
(apochromatic Twiss parameters), which will be transported through
that system without first order chromatic 
distortions.
In this paper we continue the study of the 
apochromatic properties of straight drift-quadrupole systems
and present the design of a beamline which can transport apochromatic beam ellipses 
without influence of the second and of the third order chromatic and geometric 
aberrations of the beamline transfer map.

\vspace{-0.15cm}
\section{VARIABLES, MAPS AND APOCHROMATS}

\vspace{-0.1cm}
We consider the beam dynamics in a magnetostatic system and, as usual, 
take the path length along the reference orbit $\tau$ to be the independent 
variable. We use a complete set of symplectic variables 
$\mbox{\boldmath $z$} = (x, p_x, y, p_y, \sigma, \varepsilon)^{\top}$ as 
particle coordinates.
Here $\,x,\,y\,$ measure the transverse displacements from the ideal orbit
and $\,p_x,\,p_y\,$ are transverse canonical monenta scaled with 
the constant kinetic momentum of the reference particle $p_0$.
The variables $\,\sigma\,$ and $\,\varepsilon\,$ which describe
longitudinal dynamics are 

\vspace{-0.15cm}
\noindent
\begin{eqnarray}
\sigma\,=\,c\,\beta_0 \, (t_0\,-\,t),
\hspace{0.5cm}
\varepsilon\,=\,
({\cal{E}}\,-\,{\cal{E}}_0) \,/\,(\beta_0^2 \,{\cal{E}}_0),
\label{L10_0} 
\end{eqnarray}

\vspace{-0.15cm}
\noindent
where $\,{\cal{E}}_0,\,\beta_0\,$ and $\,t_0 = t_0(\tau)\,$
are the energy of the reference particle, its velocity in terms 
of the speed of light $\,c\,$ and its arrival time 
at a certain position $\,\tau$, respectively.

We represent particle transport from the location $\tau = 0$
to the location $\tau = l$ by a symplectic map ${\cal M}$ 
and assume that the point $\mbox{\boldmath $z$} = \mbox{\boldmath $0$}$ 
is the fixed point and that the map ${\cal M}$ can be Taylor expanded 
in its neighborhood.
Additionally we assume that the transverse motion 
is dispersion free (always true for the drift-quadrupole systems) 
and uncoupled in linear approximation, which is a restriction on 
the form of the $6 \times 6$ symplectic matrix $M$ which represents
the linear part of the map ${\cal M}$.

Let $g_0$ be some function of the variables $\mbox{\boldmath $z$}$
given at the system entrance. Then its image $g_l$ at the system exit
under the action of the map ${\cal M}$ is given by the following relation

\vspace{-0.15cm}
\noindent
\begin{eqnarray}
\forall \mbox{\boldmath $z$}
\hspace{0.5cm}
g_l(\mbox{\boldmath $z$}) \,=\,
g_0({\cal M}^{-1}(\mbox{\boldmath $z$})),
\label{FP_0}
\end{eqnarray}

\vspace{-0.15cm}
\noindent
which symbolically we will write as
$\;g_l \,=\, :{\cal M}:^{-1} g_0$.

Let us consider some Courant-Snyder quadratic forms

\vspace{-0.15cm}
\noindent
\begin{eqnarray}
\left\{
\begin{array}{l}
\vspace{0.2cm}
I_x(\tau) \, = \,
\gamma_x(\tau)\, x^{2}  + 2 \alpha_x(\tau) \, x\, p_x 
+ \beta_x(\tau) \, p_x^{2}\\
I_y(\tau) \, = \,
\gamma_y(\tau)\, y^{2} \, + 2 \alpha_y(\tau) \, y\, p_y 
+ \beta_y(\tau) \, p_y^{2} 
\end{array}
\right.
\label{IFB_1}
\end{eqnarray}

\vspace{-0.15cm}
\noindent
We say that the map ${\cal M}$ is an $m$-order ($m \geq 2$) 
apochromat with respect to the incoming Courant-Snyder quadratic forms 
$I_x(0)$ and $I_y(0)$ if

\vspace{-0.15cm}
\noindent
\begin{eqnarray}
:{\cal M}:^{-1} I_{x, y}(0)
\,-\,
:M:^{-1} I_{x, y}(0) 
\,=\, 
O\left(|\mbox{\boldmath $z$}|^{m + 2}\right),
\label{IFB_3}
\end{eqnarray}
 
\vspace{-0.15cm}
\noindent
i.e. if the chromatic and
geometric distortions to the shapes of the ellipses $I_{x,y}(l)$
after the system passage come only from nonlinear map aberrations
which are of the order $m+1$ and higher.
We call the Twiss parameters that enter the
Courant-Snyder quadratic forms satisfying (\ref{IFB_3}) 
apochromatic Twiss parameters.

Up to any predefined order $m$ the aberrations of the map ${\cal M}$
can be represented through a Lie factorization as 

\vspace{-0.15cm}
\noindent
\begin{eqnarray}
:{\cal M}: \,=_m\,
\exp(:{\cal F}_{m + 1} + \ldots + {\cal F}_3:) :M:,
\label{IFB_4}
\end{eqnarray}

\vspace{-0.15cm}
\noindent
where each of the functions ${\cal F}_k$
is a homogeneous polynomial of degree $k$
in the variables $\mbox{\boldmath $z$}$ and the symbol
$=_m$ denotes equality up to order $m$ (inclusive) when
maps on both sides of (\ref{IFB_4}) are applied to the
phase space vector $\mbox{\boldmath $z$}$.

Using the representation (\ref{IFB_4}) we can state that the magnetostatic system 
is an $m$-order apochromat with respect to $I_x(0)$ and $I_y(0)$ if, and only if, 
all homogeneous polynomials ${\cal F}_k$ 
in the Lie exponential representation (\ref{IFB_4}) of the system transfer map
can be expressed as
functions of $I_x(0),\,I_y(0)$ and $\varepsilon$ only.

\vspace{-0.15cm}
\section{NONLINEAR ABERRATIONS OF STRAIGHT DRIFT-QUADRUPOLE CELL}

\vspace{-0.1cm}
The straight drift-quadrupole cell is a magnetostatic system which is 
symmetric about both, the horizontal midplane $y=0$ and the vertical 
midplane $x=0$. It means that in the Lie exponential representation of
the cell map

\vspace{-0.15cm}
\noindent
\begin{eqnarray}
:{\cal M}_c: \,=_m\,
\exp(:{\cal F}_{m+1}^c + \ldots + {\cal F}_{3}^c :) :M_c:
\label{TWO_DQ}
\end{eqnarray}

\vspace{-0.15cm}
\noindent
the polynomials ${\cal F}_{k}^c$ do not depend on the variable $\sigma$ and 
are even functions of the variables $(y, p_y)$ and of the variables $(x, p_x)$.
If the drift-quadrupole cell is not a pure drift space, then the structure of 
the polynomial ${\cal F}_{3}^c$, which is responsible for the second-order map 
aberrations, can be further clarified using the concept of the apochromatic Twiss
parameters. Let $\,\beta_{x,y}^a(\tau)$, 
$\,\alpha_{x,y}^a(\tau)\,$ and $\,\gamma_{x,y}^a(\tau)$ be the cell apochromatic 
Twiss parameters and let $\,I_{x, y}^a(\tau)\,$ be the corresponding  Courant-Snyder 
quadratic forms. Then the polynomial $\,{\cal F}_{3}^c\,$ can be written as follows

\vspace{-0.15cm}
\noindent
\begin{eqnarray}
{\cal F}_3^c
= -(\varepsilon \,/\, 2) \cdot
\big(
\xi_{x}(\beta_{x}^a) \cdot I_{x}^a(0) \,+\, 
\xi_{y}(\beta_{y}^a) \cdot I_{y}^a(0) 
\nonumber
\end{eqnarray}

\vspace{-0.4cm}
\noindent
\begin{eqnarray}
-\,l_c \cdot (\varepsilon \,/\,\gamma_0)^2
\big),
\label{MKJ_1}
\end{eqnarray}

\vspace{-0.15cm}
\noindent
where $l_c$ is the cell length, $\gamma_0$ is the Lorentz factor
of the reference particle, and $\,\xi_x(\beta_x^a)\,$ and 
$\,\xi_y(\beta_y^a)\,$ are the cell chromaticities
calculated for the apochromatic Twiss parameters 
(see more details in ~\cite{ApochromIPAC10, AchromatIPAC11}).

\vspace{-0.15cm}
\section{CONCEPTUAL SOLUTION FOR THIRD-ORDER APOCHROMAT}

\vspace{-0.1cm}
Let us consider a system constructed by a repetition of $n$ identical 
drift-quadrupole cells and let us assume that the horizontal and vertical 
transfer matrices of the $n$-cell system are equal to the plus or minus 
identity matrices, but the horizontal and vertical cell focusing matrices 
are not equal to the plus or minus identity matrices. Then, as it was shown 
in ~\cite{ApochromIPAC11}, the shape distorting effects of the second-order
aberrations of the cell map on the transport of the cell periodic
Courant-Snyder quadratic forms are washed out at the exit of the $n$-cell
system by averaging. In other words, it means that under the assumptions made
the $n$-cell system is a second-order apochromat with respect to the cell
periodic Courant-Snyder quadratic forms. And, as we will show below, the idea 
of the usage of averaging provided by repetition of $n$ identical cells for 
automatic cancellation of undesired effects can also be applied to the design 
of a third-order apochromat if the cell periodic Twiss parameters coincide 
with the cell apochromatic Twiss parameters.

So, let us start with the assumption that the cell transfer matrix allows 
periodic beam transport and that the cell periodic Twiss parameters are, 
at the same time, the cell apochromatic Twiss parameters. Then the map of the 
repetitive $n$-cell system can be expressed as follows

\vspace{-0.15cm}
\noindent
\begin{eqnarray}
:{\cal M}_{nc}: 
=_4
\exp(:{\cal S}_{3} + {\cal S}_{4} +
\tilde{\cal S}_{5} + \hat{\cal S}_{5}:) :M_c^n:,
\label{RepSingleLie_3}
\end{eqnarray}

\vspace{-0.15cm}
\noindent
where the homogeneous polynomials (aberration functions)
$\,{\cal S}_{3},\,{\cal S}_{4},\,\tilde{\cal S}_{5},\, \hat{\cal S}_{5}\,$ 
are given by the formulas

\vspace{-0.15cm}
\noindent
\begin{eqnarray}
{\cal S}_{3}(\mbox{\boldmath $z$}) =
n \cdot {\cal F}_3^c(\mbox{\boldmath $z$}),
\label{RepSingleLie_7}
\end{eqnarray}

\vspace{-0.35cm}
\noindent
\begin{eqnarray}
{\cal S}_{4}(\mbox{\boldmath $z$}) =
n \cdot  \mbox{\boldmath ${\cal R}$}({\cal F}_4^c(\mbox{\boldmath $z$})),
\;\;
\tilde{\cal S}_{5}(\mbox{\boldmath $z$}) =
n \cdot  \mbox{\boldmath ${\cal R}$}({\cal F}_5^c(\mbox{\boldmath $z$})),
\label{RepSingleLie_8}
\end{eqnarray}

\vspace{-0.35cm}
\noindent
\begin{eqnarray}
\hat{\cal S}_{5}(\mbox{\boldmath $z$}) =
(1 \,/\, 2)
\big\{
{\cal F}_3^c(\mbox{\boldmath $z$}), {\cal W}_n(\mbox{\boldmath $z$})
\big\},
\label{RepSingleLie_8_1}
\end{eqnarray}

\vspace{-0.15cm}
\noindent
the binary operation $\,\{*,*\}\,$ is the Poisson bracket,

\vspace{-0.15cm}
\noindent
\begin{eqnarray}
\mbox{\boldmath ${\cal R}$}\big(f(\mbox{\boldmath $z$})\big)
\,=\, \frac{1}{n}
\sum\limits_{m = 0}^{n - 1}
f(M_c^m \mbox{\boldmath $z$}),
\label{RepSingleLie_9}
\end{eqnarray}

\vspace{-0.35cm}
\noindent
\begin{eqnarray}
{\cal W}_n(\mbox{\boldmath $z$}) 
\,=\,
\sum\limits_{m = 1}^{[n\,/\,2]}
(n + 1 - 2 m)
\nonumber
\end{eqnarray}

\vspace{-0.35cm}
\noindent
\begin{eqnarray}
\cdot
\big(
{\cal F}_4^c(M_c^{n-m}\mbox{\boldmath $z$})
\,-\,
{\cal F}_4^c(M_c^{m-1}\mbox{\boldmath $z$})
\big),
\label{RepSingleLie_10}
\end{eqnarray}

\vspace{-0.15cm}
\noindent
and the symbol $[k]$ denotes the biggest integer which is
smaller or equal to $k$.

Our second assumption is that the periodic cell
phase advances $\,\mu_{x}^c\,$ and $\,\mu_{y}^c\,$ satisfy

\vspace{-0.15cm}
\noindent
\begin{eqnarray}
\mu_{x, y}^c \,=\, 2 \pi \,q_{x,y} \,/\, n
\;\;\;(\mbox{mod} \, 2\pi)
\label{RepCellTunes_1_1}
\end{eqnarray}

\vspace{-0.15cm}
\noindent
for some integer or half integer $q_{x}$ and $q_{y}$
which are smaller than $n$ and are such that the resonances

\vspace{-0.15cm}
\noindent
\begin{eqnarray}
2 \,\mu_{x, y}^c, 
\;\;\;
4 \,\mu_{x, y}^c, 
\;\;\;
2 \,\mu_{x}^c \,\pm\, 2\, \mu_{y}^c
\label{RepCellTunes_1_7}
\end{eqnarray}

\vspace{-0.15cm}
\noindent
are avoided. From this assumption it follows that the functions 
$\,{\cal S}_{4}\,$ and $\,\tilde{\cal S}_{5}\,$
(as a result of averaging provided by the operator $\mbox{\boldmath ${\cal R}$}$)
can be expressed as functions of the variables $\,I_{x,y}^a(0)\,$ 
and $\,\varepsilon\,$ only, and thus the $n$-cell beamline 
becomes a third-order apochromat with respect to the cell periodic
Courant-Snyder quadratic forms $\,I_{x,y}^a(0)$.

Note that the minimal number of cells required for the construction
of the third-order apochromatic beamline following the above recipe is five,
and that the first distortions to the shapes of the Courant-Snyder ellipses
$\,I_{x,y}^a(0)\,$ after the system passage are due to forth-order aberrations 
of the map $\,{\cal M}_{nc}\,$ generated by the function $\,\hat{\cal S}_{5}$.

\vspace{-0.15cm}
\section{PROOF-OF-PRINCIPLE\\ NUMERICAL EXAMPLE}

\vspace{-0.1cm}
To complete the third-order apochromat design we need a drift-quadrupole
cell with the coinciding periodic and apochromatic Twiss parameters and \
with the phase advances satisfying the conditions (\ref{RepCellTunes_1_1}) 
and (\ref{RepCellTunes_1_7}). What is the minimum number of quadrupoles 
which are required for such a cell to exist? Using thin-lens approximation 
for quadrupole focusing one can show that this number can not be smaller 
than four, and with four quadrupoles we have many examples (found by direct 
numerical search using hard-edge quadrupole models) of the cells which satisfy 
all needed requirements. The parameters of one such cell are given in the 
Table 1 and its periodic (and apochromatic) betatron functions can be seen 
in Fig.1. The horizontal and vertical phase advances of this cell are equal 
to $144^{\circ}$ and $108^{\circ}$ respectively, which means that the sequence 
of five such cells gives a third-order apochromatic beamline.

\begin{table}[h]
\vspace{-0.3cm}
\caption{
\label{tab:table1}
Parameters of apochromatic four quadrupole cell.\vspace{-0.3cm}}
\begin{center}
\begin{tabular}{|l|r||l|r|}
\hline
$k_1$       &  0.208766  &  $l_4$      &  2.211502    \\  
$k_2$       & -0.192846  &  $l_5$      &  7.825868    \\  
$k_3$       &  0.158699  &  $\beta_x$  & 15.243567    \\
$k_4$       & -0.145879  &  $\alpha_x$ & -0.464978    \\
$l_{quad}$  &  1.000000  &  $\mu_x^c$  & 144$^{\circ}$\\
$l_1$       &  6.922985  &  $\beta_y$  & 22.770092    \\
$l_2$       &  0.892888  &  $\alpha_y$ &  0.525805    \\
$l_3$       & 18.011646  &  $\mu_y^c$  & $108^{\circ}$\\
\hline
\end{tabular}
\end{center}
\vspace{-0.4cm}
\end{table}

The total number of quadrupoles in our third-order apochromatic beamline
is twenty, and it is interesting to compare its beam transfer properties 
with the beam transport through the chain of ten FODO cells which also has  
twenty quadrupoles and which is designed to have the same length and 
about the same average betatron functions. 

\begin{figure}[!htb]
    \centering
    \includegraphics*[width=70mm]{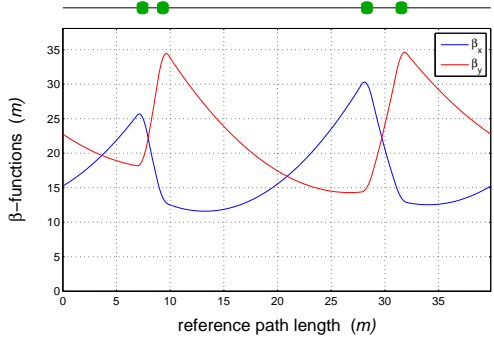}
    \vspace{-0.55cm}
    \caption{Periodic betatron functions along apochromatic 
    four quadrupole cell with $\mu_x^c = 144^{\circ}$ and $\mu_y^c = 108^{\circ}$.}
    \vspace{-0.5cm}
    \label{fig1}
\end{figure}

We chose for such comparison not one but three different FODO beamlines 
with the cell phase advances equal to $54^{\circ}$, $60^{\circ}$ 
and $72^{\circ}$ (see, for example Fig.2).
The first and the third FODO beamlines have the phase advances which are equal 
to the half of the vertical and of the horizontal phase advances of our
apochromatic four quadrupole cell respectively and, at the same time, are
second-order apochromats. The second FODO beamline is taken because, 
from one side, its phase advance $60^{\circ}$ lies in between the phase advances
of the two other beamlines, but, from the other side, its overall phase advance
is not multiple of $180^{\circ}$ and, therefore, it is not a second-order apochromat.

\vspace{-0.2cm}
\begin{figure}[!htb]
    \centering
    \includegraphics*[width=70mm]{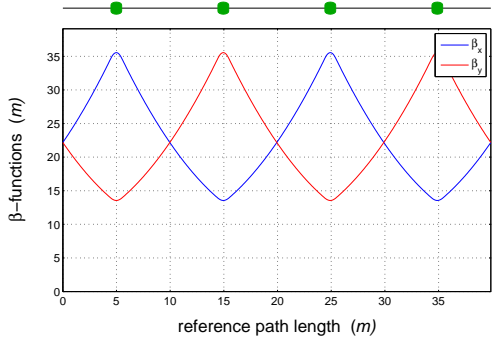}
    \vspace{-0.55cm}
    \caption{Betatron functions along two $54^{\circ}$ FODO cells.}
    \vspace{-0.2cm}
    \label{fig2}
\end{figure}

Because it is not easy to find a single quantity which will characterize 
in a clear and general way the effect of aberrations which are nonlinear 
in the transverse variables, we will restrict our comparison to the study
of the beam dynamics provided by the Hamiltonian

\vspace{-0.15cm}
\noindent
\begin{eqnarray}
H \,=\, \varepsilon \,+\,
\sqrt{(1 + \varepsilon)^2 \,-\, p_x^2 \,-\, p_y^2 \,-\, (\varepsilon \,/\, \gamma_0)^2}
\nonumber
\end{eqnarray}

\vspace{-0.4cm}
\noindent
\begin{eqnarray}
+\, (k \,/\, 2) \cdot (x^2 \,-\, y^2)
\label{Hamiltonian_1}
\end{eqnarray}

\vspace{-0.15cm}
\noindent
after its linearization with respect to $p_x^2$ and $p_y^2$,
i.e. we will look only at the effect of the energy 
deviation.\footnote{If only chromatic aberrations are of concern and the influence 
of the geometric effects can be ignored, then the minimal number of cells 
required for automatic cancellation of aberrations 
in our solution can be reduced from five to two
because in this case only the resonances $2 \mu_{x, y}^c$ 
in (\ref{RepCellTunes_1_7}) must be avoided.}

\begin{figure}[!htb]
    \centering
    \includegraphics*[width=70mm]{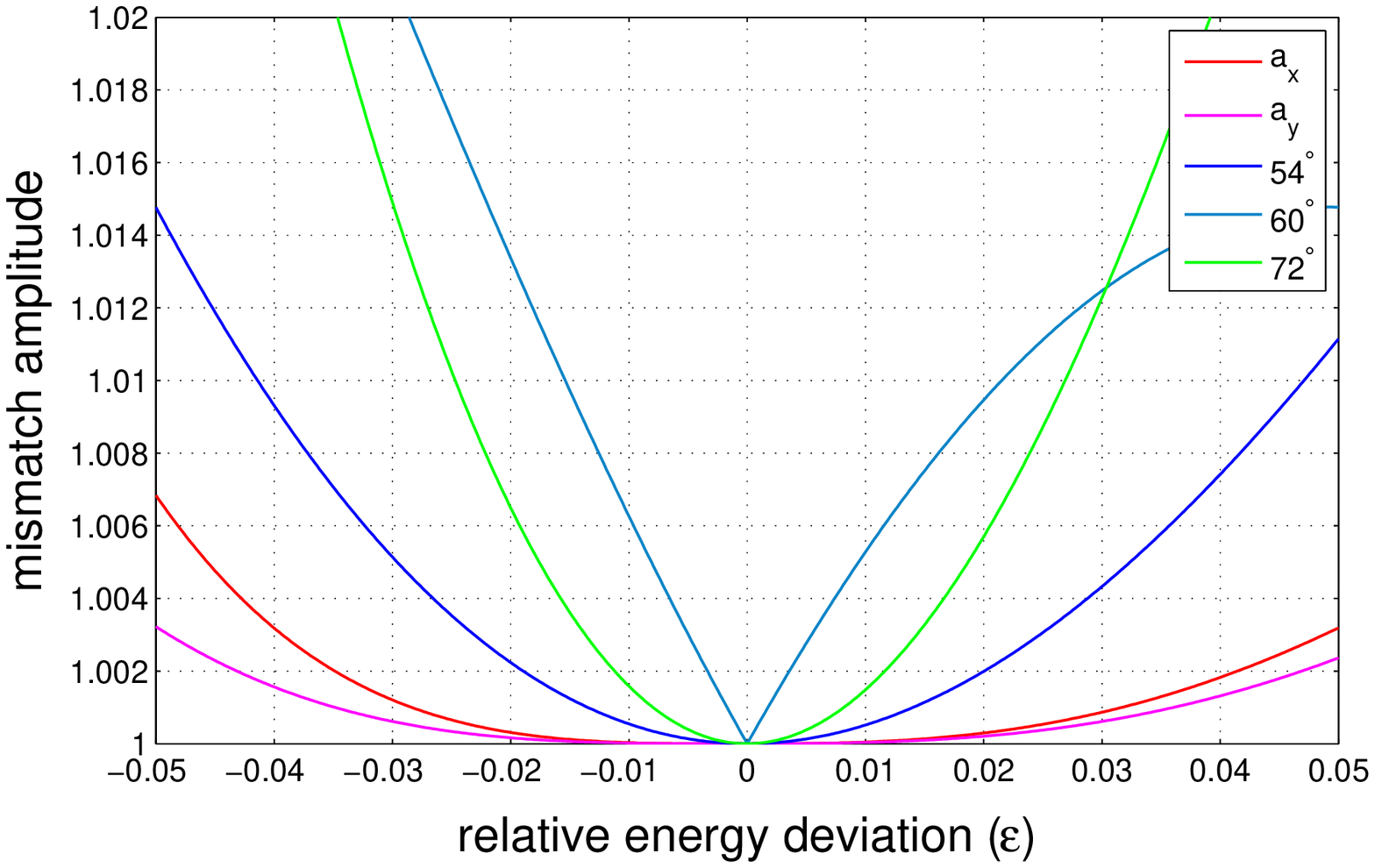}
    \vspace{-0.55cm}
    \caption{Mismatch amplitudes for different beamlines.}
    \vspace{-0.5cm}
    \label{fig3}
\end{figure}

For each beamline, for each transverse plane and for each energy offset 
we track the nominal beamline Twiss parameters (Twiss parameters which 
are periodic for the nominal energy) through the system and, at the system
exit, characterize the accumulated chromatic effects by calculating first 
the mismatch parameter $m_p$ and then the mismatch amplitude 

\vspace{-0.3cm}
\noindent
\begin{eqnarray}
a \,=\,m_p \,+\,\sqrt{m_p^2 \,-\, 1},
\label{Hamiltonian_2}
\end{eqnarray}

\vspace{-0.2cm}
\noindent
which is a measure of the created optics distortion.
The results of these calculations are presented in Fig.3, where
$a_x$ and $a_y$ denote the horizontal and vertical mismatch amplitudes
of the third-order apochromat and for each 
FODO chain only one curve is presented, because for the FODO beamlines
the horizontal and vertical mismatch amplitudes coincide.
One sees, as expected, that our third-order apochromatic beamline
demonstrates the best performance and the next comes the second-order 
apochromats with $54^{\circ}$ and $72^{\circ}$ cell phase advances.

Note that our theory is an asymptotic theory 
and uses Taylor expansion with respect to the phase space variables. 
It certainly can fail at large energy deviations, that
is demonstrated in Fig.4. 

\vspace{-0.2cm}
\begin{figure}[!htb]
    \centering
    \includegraphics*[width=70mm]{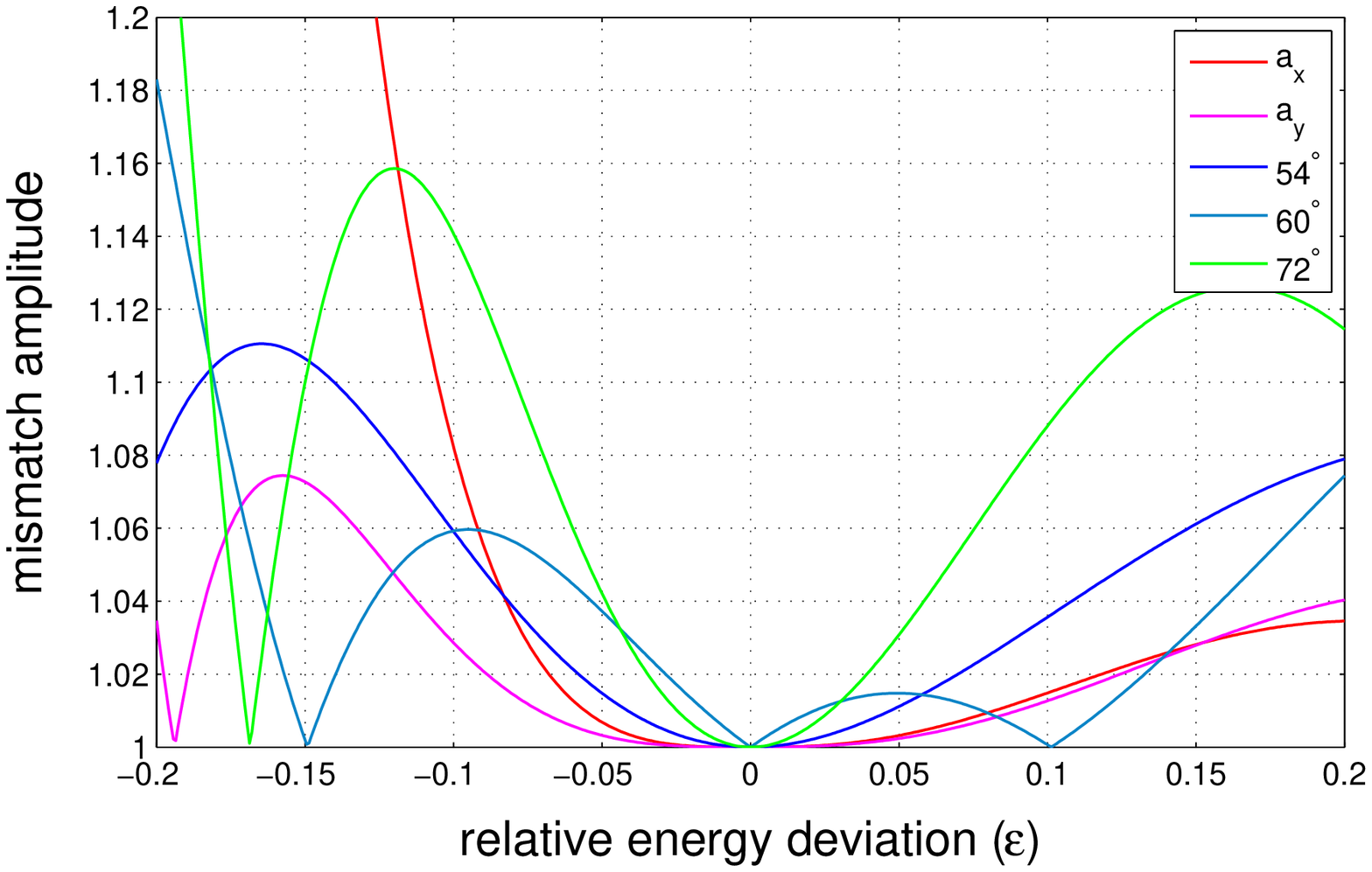}
    \vspace{-0.5cm}
    \caption{Mismatch amplitudes for different beamlines.}
    \label{fig4}
\end{figure}

\end{document}